\documentclass[twocolumn,aps,prl,10pt,superscriptaddress]{revtex4-2}
\usepackage{amsmath,amssymb,bm,graphicx,hyperref}

\begin{document}

\title{Misinformation Dynamics in Social Networks
}

\author{Jeff Murugan}
\affiliation{The Laboratory for Quantum Gravity \& Strings and,\\
Department of Mathematics and Applied Mathematics,
Univeristy of Cape Town,Private Bag, Rondebosch 7700, South Africa}

\begin{abstract}
\noindent
Information transmitted across modern communication platforms is degraded not only by intentional manipulation (disinformation) but also by intrinsic cognitive decay and topology-dependent social averaging (misinformation). We develop a continuous-fidelity field theory on multiplex networks with distinct layers representing private chats, group interactions, and broadcast channels. Our analytic solutions reveal three universal mechanisms controlling information quality: (i) groupthink blending, where dense group coupling drives fidelity to the initial group mean; (ii) bridge-node bottlenecks, where cross-community flow produces irreversible dilution; and (iii) a network-wide fidelity landscape set by a competition between broadcast truth-injection and structural degradation pathways. These results demonstrate that connectivity can reduce information integrity and establish quantitative control strategies to enhance fidelity in large-scale communication systems.
\end{abstract}

\maketitle

\section{Introduction}
\noindent
There is an old adage that says: ``A lie can travel half way around the world while the truth is putting on its shoes", meaning that a lie can spread quickly and widely, while the truth takes more time to catch up and be believed. From public health to political discourse, the integrity of information is a cornerstone of modern society. Yet, a defining feature of our digital age is the rapid degradation of information fidelity as it propagates through social networks. While traditional models of information spread often draw an analogy to biological epidemics \cite{RevModPhys.87.925,doi:10.1073/pnas.2102141118}, this framework captures reach but neglects a critical dimension: the progressive distortion and loss of nuance that transforms accurate information into misinformation. Understanding the dynamics of information decay—not merely its spread—is therefore a fundamental challenge with profound implications for the health of our digital ecosystems.\\

\noindent
The prevailing paradigm in network science models information cascades using compartmental models such as the susceptible-infectious-recovered (SIR) models made famous during the Covid-19 pandemic \cite{RevModPhys.81.591,porter2016dynamical}. This approach, while powerful, is insufficient for describing how complex information evolves. Recent work has begun to explore social influence and opinion dynamics \cite{article,PhysRevE.88.050801}, but often treats these as shifts along a continuous spectrum, without explicitly linking the rate of distortion to the underlying multi-layered topology of human interaction. Real-world social platforms are not single-layer networks; they are multiplex structures comprising distinct communication channels—such as private chats, group forums, and broadcast channels—each with characteristic speed, trust, and social pressure. A unified theory explaining how this layered topology governs the very fidelity of information remains an open problem.\\

\noindent
In this Letter, we introduce a dynamical model of information fidelity on a multiplex network to bridge this gap. We define a fidelity field $F_{i}(t)\in[0,1]$ for each node $i$, representing the accuracy of its held information. The dynamics are governed by layer-specific mechanisms:
\begin{itemize}
    \item {\it Private layers} facilitate slow, high-fidelity diffusion via pairwise contacts,
    \item {\it Group layers} induce rapid but often destructive consensus averaging, leading to a groupthink-driven loss of fidelity, and
    \item {\it Broadcast layers} provide one-way fidelity injection from authoritative sources.
\end{itemize}

\noindent
We will demonstrate that the interplay between these layers results in a universal degradation physics. Analytically, we derive the existence of a sharp fidelity drop upon information entry into group structures and identify a critical group size for distortion. We show that bridge nodes, which interconnect groups, act as fidelity bottlenecks whose integrative capacity controls cross-network contamination. Finally, we present a closed-form expression for the network's steady-state fidelity landscape, revealing it is a function of topological parameters rather than initial conditions alone. Our work establishes a quantitative link between network architecture and information integrity, offering a new lens through which to design more robust communication systems.

\section{Model and Framework}
\noindent
To quantify the degradation and transmission of information, we move beyond binary-state contagion models and introduce a continuous fidelity field
$F_i(t)\in[0,1]$,
defined on each node $i$ of a multiplex network.  $F_i=1$ denotes perfect preservation of the original message, while $F_i=0$ corresponds to complete distortion.  The time-evolution of $F_i$ reflects the interplay between intrinsic cognitive decay and the layer-specific coupling between users. We consider a multiplex network composed of three communication layers (see Fig.1): a {\it private} layer $L_1$, representing sparse one-to-one exchanges (undirected edges); a {\it group} layer $L_2$, consisting of dense, all-to-all cliques that mimic group chats; and
a {\it broadcast} layer $L_3$, a directed graph from broadcaster nodes to their followers.

\begin{figure}[h]
    \centering
    \includegraphics[width=0.8\linewidth]{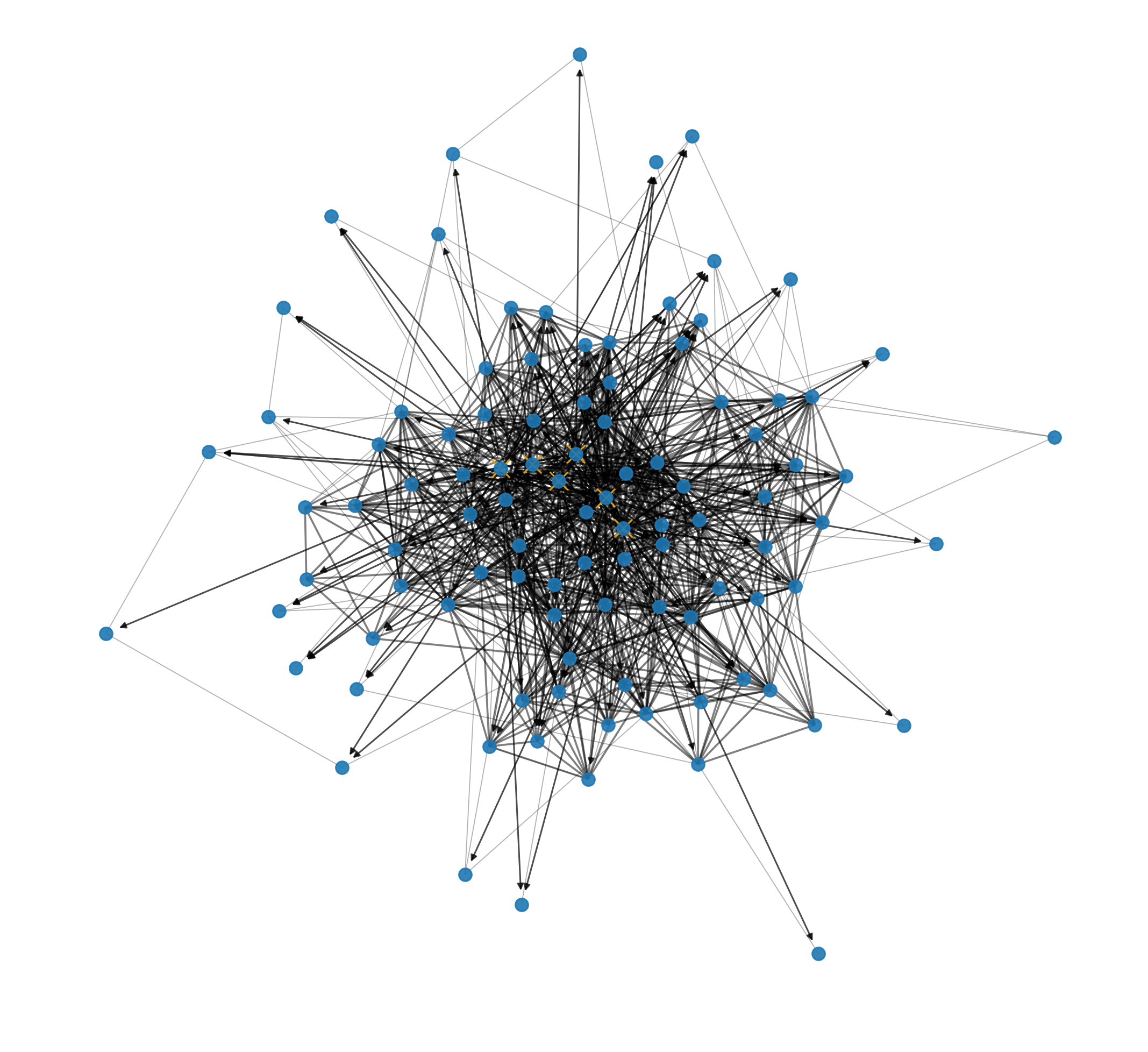}
    \caption{Multiplex topology of a WhatsApp-like network. Nodes represent users and edges represent three channel types: private 1-to-1 chats (thin undirected edges), group cliques (dense patches), and directed broadcast connections (arrows). Black-rimmed nodes are broadcasters. This structure defines the heterogeneous coupling operators in Eq. \eqref{fid_dyn}.}
    \label{fig:Lambda_param}
\end{figure}
\noindent
We conjecture that the evolution of information fidelity is governed by the dynamical system
\begin{eqnarray}\label{fid_dyn}
    \frac{dF_i}{dt}
=-\delta F_i
-\beta F_i^{2}
+\sum_{\ell=1}^{3}\Gamma_{\ell}\,D_i^{(\ell)}[\{F_j\}],
\end{eqnarray}
where the first two terms describe intrinsic degradation and the summation encodes social diffusion across layers \cite{PhysRevLett.110.028701}. The decay terms have a clear physical meaning \cite{6773024}:
\begin{itemize}
    \item The linear forgetting term $-\delta F_i$ is the simplest possible fidelity-loss process consistent with a finite memory lifetime; a user gradually forgets details of the information they hold. In the absence of social reinforcement then, fidelity decays exponentially as $F_i(t)=F_i(0)e^{-\delta t}$, with decay rate $\delta$, analogous to the relaxation of a memory state in cognitive psychology or a signal mode in a noisy physical channel.
    \item However, forgetting is not neutral with respect to the richness of the message. Detailed, nuanced content is lost disproportionately quickly, consistent with the fact that compression leads to irrecoverable loss of information content. This is the motivation behind the nonlinear “complexity decay’’ term $-\beta F_i^2$, which ensures that fidelity vanishes more rapidly when the message contains more fine-grained structure.
\end{itemize}

\noindent
Each operator $D_i^{(\ell)}$ in the sum implements a distinct mode of social coupling with strength $\Gamma_\ell$, Specifically,
\begin{enumerate}
    \item Private-chat diffusion ($\ell=1$):
    $D_i^{(1)}=\!\!\sum_{j\in N_1(i)}(F_j-F_i)$,
    where $N_1(i)$ are private-layer neighbors.  This term describes slow, high-fidelity pairwise refinement through trusted interactions,
    \item Group-consensus diffusion ($\ell=2$):
    $D_i^{(2)}=\!\!\sum_{g\in G(i)}(\bar F_g-F_i)$,
    where
    $\bar F_g=\frac{1}{|g|}\!\sum_{j\in g}F_j$.
    Here $G(i)$ lists all the groups containing node $i$.  This operator drives rapid alignment to the local group mean, capturing the reductive “groupthink’’ dynamics of collective discussions.
    \item Broadcast injection ($\ell=3$):
    $D_i^{(3)}=\!\!\sum_{b\in B(i)}(F_b-F_i)$,
    with $B(i)$ the set of broadcasters followed by the user at node $i$.  This term represents one-way fidelity inflow from authoritative sources, with no reciprocal feedback.
\end{enumerate}

\noindent
The intrinsic coefficients $\delta,\beta$ control the internal, content-dependent decay, while ($D_1,D_2,D_3$) or equivalently the couplings ($\Gamma_1,\Gamma_2,\Gamma_3$) quantify the topology-dependent spreading rates in the three communication channels.  The non-linear consensus operator $D^{(2)}$ is a new ingredient. It provides a minimal mathematical realization of “groupthink’’, a mechanism that is, as far as we are aware, absent from standard epidemic or opinion-dynamics models. Together, these define a compact field theory of information fidelity that couples memory loss, content complexity, and multi-layer social dynamics.  This framework permits analytic derivation of macroscopic degradation laws and critical thresholds directly from the microscopic interaction topology.\\

\section{Fidelity dynamics}
The dynamics of Eq. \eqref{fid_dyn} admit several analytical reductions that expose universal mechanisms of information degradation. Three representative results illustrate the essential physics; a groupthink-induced fidelity drop, a bridge-node bottleneck, and an emergence of a network-wide fidelity landscape. The densest layer, composed of group cliques, is the principal site of distortion. In a group $g$ of size $m$, each member obeys
\begin{eqnarray}\label{group_dyn}
    \dot F_i=D_2\!\!\sum_{j\in g,\,j\neq i}\!(F_j-F_i)-\delta F_i\,,
\end{eqnarray}
with average fidelity $\bar F_g=m^{-1}\sum_j F_j$.
Equation \eqref{group_dyn} can be recast as
\begin{eqnarray}
    \dot F_i = mD_2(\bar F_g - F_i) - \delta F_i\,.
\end{eqnarray}
Summing over all nodes yields $\dot{\bar F}_g = -\delta \bar F_g$, demonstrating that social coupling cancels exactly in the group average. The mean fidelity decays only through intrinsic forgetting.  The coupling $D_2$ only enforces rapid convergence of individuals toward this decaying mean.
In the strong-coupling regime ($D_2\!\gg\!\delta$),
$F_i(t)\!\to\!\bar F_g(0)e^{-\delta t}+O(e^{-mD_2t})$.
The group acts as an information blender in the sense that the final fidelity of each member is fixed by the initial group average.  A single low-fidelity participant ($F_j(0)\!\ll\!1$) can irreversibly contaminate the collective state giving a quantitative expression of groupthink.\\
\begin{figure}[h]
    \centering
    \includegraphics[width=0.9\linewidth]{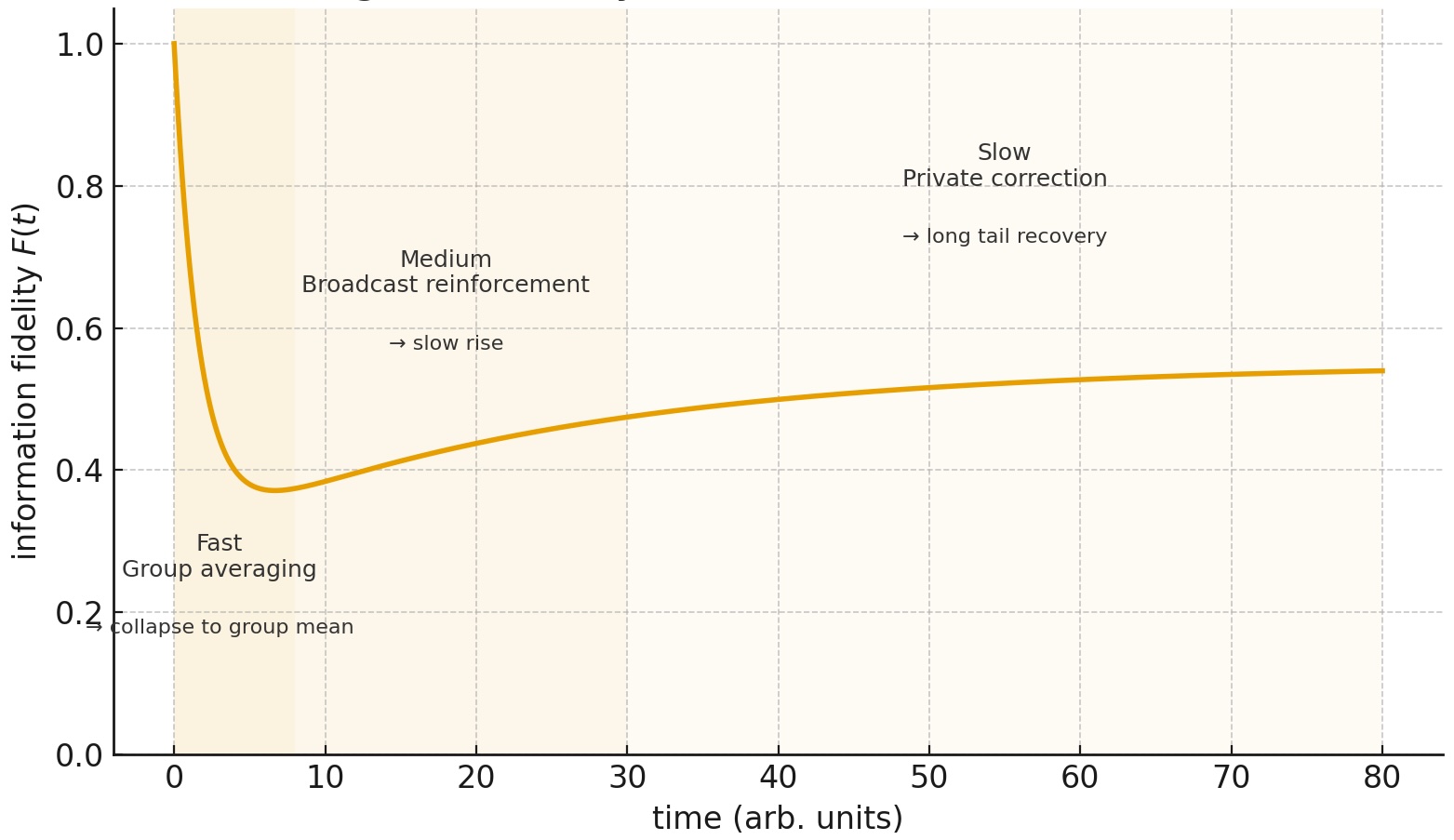}
    \caption{Characteristic fidelity decay curve.
Rapid initial drop reflects fast group-blending dynamics, governed by $mD_2$; slow recovery arises from broadcast-driven reinforcement $D_3$.}
    \label{fig:Char_curve}
\end{figure}

\noindent
Nodes connecting multiple groups dominate cross-layer information flow. The fidelity of a bridge node $B$ attached to $k$ groups evolves as
\begin{eqnarray}
    \dot F_B = \sum_{j=1}^{k} m_j D_2(\bar F_{g_j}-F_B)-\delta F_B,
\end{eqnarray}
where $m_j$ and $\bar F_{g_j}$ are the size and mean fidelity of group $g_j$. Assuming similar group sizes ($m_j\simeq m$), the steady state satisfies
$F_B^*(kmD_2+\delta)=mD_2\sum_{j=1}^{k}\bar F_{g_j}$,
or equivalently $F_B^*=(D_2\sum_{j=1}^{k}\bar F_{g_j})/
(kD_2+\delta/m)$. $F_B^*$ is a weighted average of the source fidelities, but the denominator term $kD_2$ introduces an integrative load.  As a node bridges more groups, its steady-state fidelity decreases, forming a bottleneck that propagates distortion between communities.  Even a single low-fidelity source reduces the fidelity of all connected groups—an unavoidable topological constraint on information preservation.\\

\noindent
A mean-field approximation reveals the macroscopic steady state. Let $\langle F\rangle$ denote the population-averaged fidelity and $p_b$ the fraction of nodes connected to high-fidelity broadcasters ($F\!\approx\!1$). The broadcast layer contributes a fidelity source term $D_3p_b(1-\langle F\rangle)$, while private interactions homogenize fluctuations without altering the mean. Group and bridge interactions supply a degradation sink proportional to
$\langle k_{\mathrm{bridge}}\rangle f(\langle m\rangle)\langle F\rangle$, where $f(\langle m\rangle)$ increases with group size. Balancing gain and loss yields
\begin{eqnarray}\label{bal-loss-gain}
    D_3p_b(1-\langle F\rangle)
-\delta\langle F\rangle
-\xi\langle k_{\mathrm{bridge}}\rangle f(\langle m\rangle)\langle F\rangle=0,
\end{eqnarray}
where $\xi$ is the effective distortion coefficient per bridge node that quantifies the cross-group contamination efficiency of misinformation. Equation \eqref{bal-loss-gain} has the steady-state solution
\begin{eqnarray}\label{steady-sol}
    \langle F\rangle
    =\frac{D_3p_b}
    {D_3p_b+\delta+\xi\langle k_{\mathrm{bridge}}\rangle f(\langle m\rangle)}.
\end{eqnarray}

\noindent
Eq. \eqref{steady-sol} quantifies a competition between a fidelity source (broadcast truth injection) and two classes of fidelity sinks: intrinsic forgetting ($\delta$) and topology-induced degradation ($\xi\langle k_{\mathrm{bridge}}\rangle f(\langle m\rangle)$). Hence the overall information quality depends not on connectivity {\it per se}, but on the balance between corrective and contaminative pathways—too many bridges or large groups render the network intrinsically “leaky’’. The hierarchy of couplings produces three distinct dynamical regimes:
\begin{eqnarray}\label{dynamical_regime}
    \tau_{\mathrm{grp}}\!\sim\!(mD_2)^{-1},
    \quad
    \tau_{\mathrm{bdc}}\!\sim\!D_3^{-1},
    \quad
    \tau_{\mathrm{pvt}}\!\sim\!\frac{L^2}{D_1\langle k_1\rangle}.
\end{eqnarray}

\noindent
Group equilibration is rapid, broadcast influence operates on intermediate timescales, and private correction is slow and diffusive.  This separation naturally explains the characteristic fidelity-decay curve: a fast initial drop due to group blending, followed by a slow recovery driven by pairwise refinement.
\begin{figure}[h]
    \centering   \includegraphics[width=0.9\linewidth]{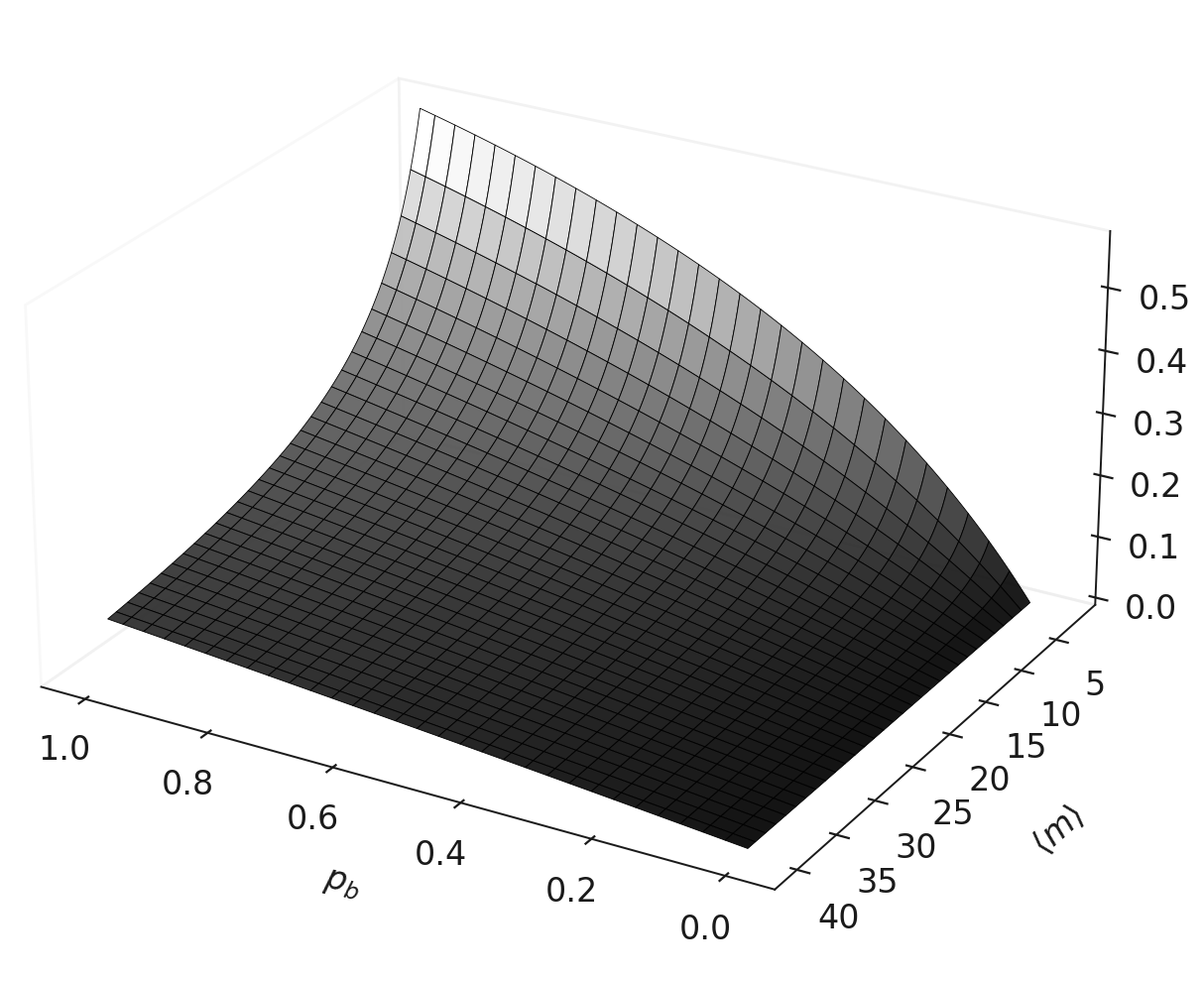}
    \caption{Steady-state average fidelity $\langle F^*\rangle$  is shown as a function of broadcast coverage $p_b$, average group size $\langle m\rangle$, and average bridge connectivity $\langle k_{\mathrm{bridge}}\rangle$. Increased intergroup connectivity and large group size drive fidelity loss.}
    \label{fig:Fidelity_land}
\end{figure}

\noindent
The competition between the characteristic timescales in Eq. \eqref{dynamical_regime} gives rise to distinct structural regimes of information fidelity, which are mapped in the ($\langle m\rangle, D_3$) plane in Fig. 4. In the red region, group consensus dynamics act faster than both broadcast reinforcement and private correction ($\tau_g < \tau_b,\,\tau_g < \tau_p$), driving a rapid collapse of fidelity toward the initial group average—a topology-induced “groupthink” phase in which distortion becomes self-amplifying. The dashed boundary ($\tau_b = \tau_p$) marks the operational stability threshold: only above this line can authoritative broadcast signals intervene on a timescale short enough to prevent collapse. The solid line ($\tau_g = \tau_b$) is subdominant in this regime, explaining why increased broadcast strength must simultaneously overcome both intra-group mixing and limited private connectivity to restore fidelity. Thus, the phase diagram reveals a structural law: platforms with large groups and insufficient broadcast reach inherently favor misinformation proliferation, whereas stronger broadcast coupling stabilizes information fidelity against topological degradation.

\begin{figure}[h]
    \centering   \includegraphics[width=0.9\linewidth]{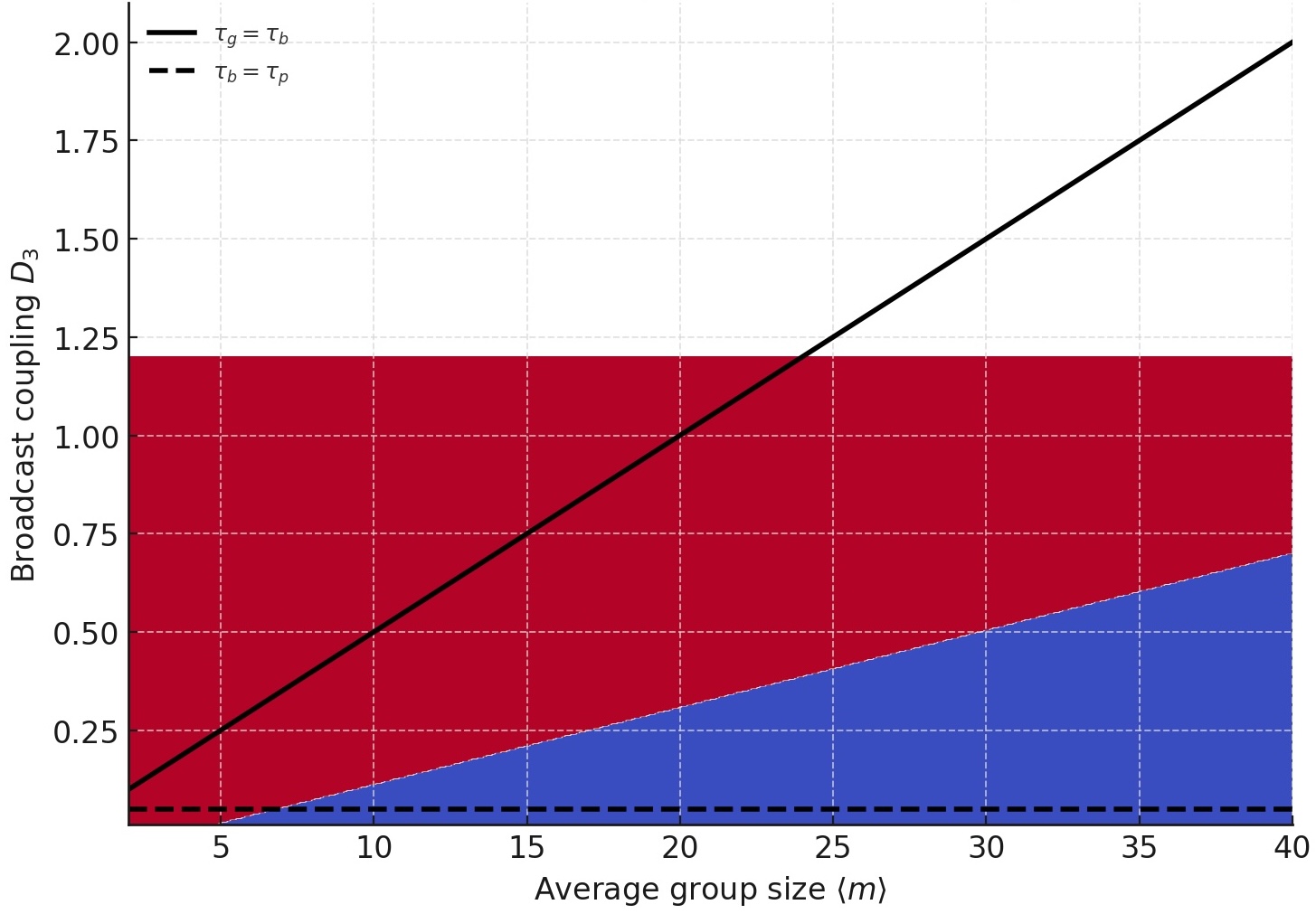}
    \caption{Structural phase diagram showing regimes of information fidelity as a function of average group size $\langle m\rangle$ and broadcast strength $D_3$. In the red region, rapid group averaging overwhelms broadcast and private correction ($\tau_g < \tau_b, \tau_g < \tau_p$), leading to a groupthink-driven fidelity collapse. In the blue region, broadcast reinforcement arrives soon enough to prevent collapse. The dashed boundary ($\tau_b=\tau_p$) defines the operational stability threshold from Eq. \eqref{dynamical_regime}.}
    \label{fig:Phase diag}
\end{figure}
\section{Discussion and implications}
\noindent
The analysis presented above establishes that information quality in multiplex social networks is governed by a competition between distortion and reinforcement, which emerges from the interplay of intrinsic cognitive decay and heterogeneous social coupling.  Three universal mechanisms arise.
First, dense group interactions lead to groupthink blending, where rapid equilibration drives individual fidelity to the initial group mean regardless of coupling strength.  This mathematically explains a familiar sociological effect: collective deliberation can suppress individual accuracy when even a single participant introduces noise.
Second, bridge nodes — individuals embedded in multiple groups — generate structural bottlenecks where information from distinct communities is mixed and diluted.  These nodes serve as unavoidable entropy sources: greater inter-group connectivity increases the network’s vulnerability to degradation, even when all but one community holds accurate beliefs.
Third, a global fidelity landscape emerges from a balance between (i) truth-injection through directed broadcast channels and (ii) topology-induced sinks driven by group and bridging structure.  This yields a closed-form expression for the population-level fidelity (Eq. \eqref{steady-sol}) that isolates the core control parameters: fraction of nodes receiving high-fidelity input, average group size, and bridge connectivity.\\

\noindent
These mechanisms have concrete implications for the design and governance of digital communication systems.  They demonstrate that network connectivity is not intrinsically beneficial for information integrity.  Rather, connectivity that disproportionately amplifies group blending and bridge mixing can push the system into a low-fidelity phase even when intrinsic forgetting is weak and broadcast accuracy is high.  Conversely, distributed access to high-fidelity broadcast channels increases the source term $D_3p_b$, raising the global fixed point $\langle F^*\rangle$.\\

\noindent
Our framework highlights a fundamental vulnerability of WhatsApp-like communication platforms: private moderation and fact-checking arrive on the slowest timescale ($\tau_{\mathrm{pvt}}$), while distortion spreads rapidly through group interactions ($\tau_{\mathrm{grp}}$).  This inherent mismatch in timescales makes reactive correction insufficient, pointing instead toward structural interventions such as controlling group size, reducing excessive bridging, or increasing the direct reach of trusted broadcast channels. Importantly, our results are parameter-robust and do not depend on particular cognitive or behavioral assumptions.  Consequently, we conjecture that any system in which information fidelity is locally averaged in clusters and globally transported by bridge nodes will exhibit similar macroscopic degradation patterns.  Thus, the physics derived here constitutes a universal theory of information fidelity in social communication networks, with immediate relevance to misinformation mitigation, platform architecture, and public policy.

\section{Acknowledgements}
\noindent
We would like to thank Mughtar Parker for a discussion that inspired this work, Georgie Roussos whose unpublished project on vaccine distribution in layered networks inspired our mathematical model, and especially Amanda Weltman for clarifying aspects of the network structure and a critical reading of the manuscript.

\bibliography{bib}

\end{document}